\documentstyle[12pt]{article}
\begin{document}

\newcommand{\bc}{\begin{center}}
\newcommand{\ec}{\end{center}}
\newcommand{\eq}{\begin{equation}}
\newcommand{\en}{\end{equation}}
\newcommand{\qq}{$q\bar q$~}

\begin{titlepage}
\begin{flushright}
 hep-ph/9605320
\end{flushright}
\vspace*{1cm}

\bc
{\Large The Dynamical Retardation Corrections to the Mass Spectrum of Heavy Quarkonia}
\ec

\vspace{.6cm}
\bc
{\bf T.Kopaleishvili$^{\dagger}$, A.Rusetsky$^{\ddagger}$}
\ec

\bc
{\it High Energy Physics Institute, Tbilisi State University}
\ec
\bc
{\it 9, University st., 380086, Tbilisi, Republic of Georgia}
\ec

$^{\dagger}$ E-mail: root@hepitu.acnet.ge, root@presid.acnet.ge

$^{\ddagger}$ E-mail: george@mdc.acnet.ge

\vspace{.9cm}
\bc
{\bf Abstract}
\ec
\vspace{.3cm}

A new version for the relativistic generalization of the wide class of
static quark--antiquark confining potentials is suggested. The comparison of
this approach with other ones, known in literature, is considered.
With the use of Logunov--Tavkhelidze quasipotential approach the first--order
retardation corrections to the heavy quarkonia mass spectrum are calculated.
As expected, these corrections turn out to be small for all low--lying heavy
meson states.

\vspace{5cm}

\begin{flushbottom}

\end{flushbottom}

\end{titlepage}
\section{Introduction}

\par
Within the framework of the constituent quark model it is quite natural to
study the properties of bound \qq systems on the basis of Bethe--Salpeter
equation. Despite the remarkable success in the quantitative description
of the bound--state masses and formfactors, achieved with the use of
instantaneous (static) interaction kernel in this equation [1-4], the
completely relativistic approach to the problem is still lacking. From the
physical point of view one can expect that the dynamical retardation corrections
(i.e. the corrections, coming from the explicit dependence of the quark--quark
interaction kernel on the relative energy variables) to the bound--state
characteristics must be small for heavy quarkonia and may become significant in the
light--quark sector. Moreover, one expects smooth static limit, when the
masses of the constituent quarks tend to the infinity. In practice, however,
the situation is more complicated, owing to the infrared--singular behaviour
of the (phenomenological) confining quark--quark interaction kernels.
Since at the present stage the exact derivation of such relativistic
kernels directly from QCD is not known, the different prescriptions
are assumed for the {\it ad hoc} relativistic generalization [5-11]
of the static interquark confining potentials. As to the one--gluon
exchange part of the interquark potential, which has proven to be
significant in the quantitative description of meson data, it can be uniquely
generalized to 4 dimensions with the use of the field--theoretical arguments.
In difference with the one--gluon exchange potential, the "relativization"
of the confining potentials, in general, introduces a new mass parameter
in the theory \cite{LUCHA,BRAM,KR,KRYAF}, which may be fixed, using the additional
constraints either directly on the relativistic counterpart of this potential
\cite{LUCHA,BRAM,KR,KRYAF} or on the bound--state equation \cite{GROSS}. Neither
of these constraints can be preferred from the physical point of view,
rendering the identification of the dynamical retardation effect ambiguous.

The existence of the additional free parameter in the theory stems from the necessity
of the infrared regularization of the "confining" kernel in 4 dimensions.
As a result, the smooth static limit, in general, does not exist \cite{BRAM,KRYAF}
and the dynamical retardation corrections to the bound--state characteristics turn
out to be large in the heavy--quark sector \cite{BRAM,KRYAF}, rendering doubtful even
the concept of the confining interaction in 4 dimensions \cite{BRAM}. Moreover,
the straightforward relativistic generalizations for some widely used confining potentials
(e.g. for the oscillator potential) seem do not possess the discrete
spectrum \cite{BIS} in the entire energy region \cite{KRYAF}. Thus, the further
investigation of the problem in the framework of Bethe--Salpeter equation
with the relativistic kernels is needed in order to clarify the notion of
the confining force in 4 dimensions. In other words, one has to answer the
question, whether there exist relativistic kernels which are compatible
with confinement, lead to a smooth static limit and, therefore, can be expected to be
obtained from QCD with the use of nonperturbative methods.

\section{The Relativistic Generalization of the Static Confining Potentials}

\par
In the present work we suggest a new prescription for the relativistic
generalization of the wide class of static confining potentials, where
some difficulties which are inherent for the prescriptions, suggested
earlier [5-10], are presumably avoided. In order to clarify the physical
meaning of this prescription, let us consider the case of two space--time dimensions.
We start from the power--law potentials $V_{\alpha}(r)=r^{\alpha}$. The
corresponding Bethe--Salpeter kernel in two dimensions has the form:
\eq
K_{\alpha}(x)=K_{\alpha}(x_0,x_1)=\delta (x_0)V_{\alpha}(|x_1|)=
\delta (nx)(-x^2)^{\alpha /2}
\en
\noindent
where $x^2=x_0^2-x_1^2$ and $n=(1,0)$. We assume that the relativistic
generalization of this expression is the "Lorentz--average" over all possible
directions of the unit vector $n$ which can be obtained from the vector $(1,0)$
by the proper Lorentz transformations. In 2 space--time dimensions
this vector transforms as $n_0'(\varphi)=n_0\cosh\varphi-n_1\sinh\varphi$,
$n_1'(\varphi)=n_1\cosh\varphi-n_0\sinh\varphi$, $\varphi=\frac{v}{c}$ and
$-\infty<\varphi<\infty$ and the group--invariant measure is $d\varphi$.
Then the "average" over all possible directions of $n$ can be defined as follows:
\eq
\bar K_{\alpha}(x)=(-x^2)^{\alpha /2}<\delta (nx)>_n=
(-x^2)^{\alpha /2}\frac{1}{{\cal N}}\int_{-\infty}^{\infty} d\varphi
\delta (x_0n_0'(\varphi)-x_1n_1'(\varphi))
\en
\noindent
where ${\cal N}$ is an arbitrary Lorentz--invariant normalization constant.
The integral in the r.h.s. of eq. (2) can be easily evaluated:
\eq
\int_{-\infty}^{\infty} d\varphi \delta(x_0n_0'(\varphi)-x_1n_1'(\varphi))=
\theta(-x^2)(-x^2)^{-1/2}
\en

The constant ${\cal N}$ is fixed from the condition that in the static
limit the kernel (2) reduces to its static counterpart $V_{\alpha}(r)$:
$\int_{-\infty}^{\infty} dx_0\bar K_{\alpha}(x_0,x_1)=V_{\alpha}(|x_1|)$.
This gives:
\eq
\bar K_{\alpha}(x)=\frac{\Gamma(1+\frac{\alpha}{2})}{\sqrt{\pi}\Gamma(\frac{1}{2}+\frac{\alpha}{2})}
\theta (-x^2)(-x^2)^{(\alpha -1)/2}
\en

Thus, the relativistic kernel (4) in two dimensions can be interpreted merely
as the static potential, averaged over all possible directions of the quantization
axe, which are equally acceptable from the physical point of view.
For the case of 4 space--time dimensions the straightforward evaluation
of the integral over the group--invariant measure in analogy with the
eqs. (2)--(3) is not possible. Despite this the "average" of the static
kernel (1) over all possible directions of the unit vector $n$ can be defined
for this case as well, and we again arrive at the expression (4)
(for the details see Appendix A).
Note that the Coulombic interaction in 4 dimensions ($\alpha =-1$) is excluded
in the expression (4).

Having obtained the expression (4), we can forget about its "derivation"
and consider (4) merely as an {\it anzats} for the relativistic generalization
of the power--law potentials $V_{\alpha}(r)=r^{\alpha}$.
Further, we consider the exponential static potential $V_{exp}(r)=e^{-\mu r}$.
Expanding this potential in powers of $r$ and using eq. (4) for each term of this
sum, we obtain the prescription:
\eq
e^{-\mu r}\rightarrow\theta(-x^2)\left(-\frac{\mu}{2}J_0(\mu\sqrt{-x^2})+
\frac{1}{\pi}(-x^2)^{-1/2}~_1F_2(1;\frac{1}{2},\frac{1}{2};-\frac{\mu^2x^2}{4})\right)
\en
\noindent
where $J_{\nu}$ and $_pF_q$ denote, respectively, the Bessel and hypergeometric
functions. Eq. (5) enables one to apply the procedure of the relativistic
generalization to the wide class of potentials which can be written in the
following form:
\eq
V(r)=\int d\mu C(\mu)e^{-\mu r}
\en
\noindent
where $C(\mu)$ must obey to the certain conditions in order to provide the convergence
of the integral over $d\mu$ in the relativistic case.

Now we discuss the particular case of linear confinement ($\alpha =1$) in detail.
Note that the anzats similar to (4) for the gluon propagator was employed
earlier in ref. \cite{CORN,BLAHA}. The Fourier transform of eq. (4) for $\alpha =1$
reads:
\eq
\bar K_1(q)=-4\pi\left(\frac{1}{(q_0^2-(|{\vec q}|+i0)^2)^2}+
		       \frac{1}{(q_0^2-(|{\vec q}|-i0)^2)^2}\right)
\en
\noindent
i.e., in other words, in order to obtain the kernel (7) in the ladder
approximation, the principal--value prescription must be used in the
gluon propagator instead of the familiar causal one. One can find here
a close analogy with the 2--dimensional QCD (see, e.g. [14-16]), where
the gluon exchange is definitely known to confine quarks as well as with
the gluon propagator from ref. \cite{CHENGTSAI}. As it is well known
\cite{BLAHA}, due to the
principal--value prescription in (7) the most severe infrared singularities
in the equations for the Green's functions are avoided. To demonstrate this,
we consider the Schwinger--Dyson equation for the quark propagator written in the
Feynman gauge:
$$
S^{-1}(p)=S_0^{-1}(p)+ig^2C_F\int\frac{d^4q}{(2\pi)^4}
\gamma_{\mu}S(p-q)\Gamma^{\mu}(p-q,p)D(q)=
$$
$$
=S_0^{-1}(p)+ig^2C_F\int\frac{d^4q}{(2\pi)^4}
\gamma_{\mu}\left( S(p-q)\Gamma^{\mu}(p-q,p)-S(p)\Gamma^{\mu}(p,p)\right) D(q)+
$$
\eq
+ig^2C_F\gamma_{\mu}S(p)\Gamma^{\mu}(p,p)\int\frac{d^4q}{(2\pi)^4}D(q)
\en
\noindent
where $S(p)$ and $S_0(p)$ are full and free fermion propagators, $\Gamma^{\mu}(k,p)$
is the dressed quark--gluon vertex function, $D(q)$ is the dressed gluon propagator,
$g$ is the quark--gluon coupling constant and $C_F$ is the quadratic Casimir
operator in the fundamental representation. If the principal--value anzats
(7) is used for $D(q)$, then the last term in eq. (8) is finite and we
obtain the subtracted form of the Schwinger--Dyson equation, where the
infrared singularity is softened. In difference with this, in the Euclidean
formulation of Schwinger--Dyson equations, which implies the causal prescription
for all propagators after the continuation to the Minkowski space, the
last term in eq. (8) diverges and requires the infrared regularization
\cite{BADRI}. Consequently, the relativistic generalization of the static interquark
interaction kernels introduces an additional mass parameter when the causal prescription
is used in these kernels or, equivalently, when one works in the Euclidean
space from the beginning. However, no such parameter is needed, when one uses
the principal--value prescription (7) or, in general, the prescriptions (4)--(6).
In the absence of the additional free mass parameter one expects the existence
of the smooth static limit for the calculated meson observables, while the corrections
to the static limit should allow for the unambiguous evaluation.

\section{First--Order Quasipotential for the \qq Systems}

\par
The formulated approach for the construction of the relativistic confining kernels
below we apply to the calculation of the dynamical retardation corrections
to the bound \qq system masses. The static confining potential was taken to
be linear + constant term: $V_c(r)=kr+c$ and the spin structure was chosen to be
the equal--weight mixture of scalar and the fourth component of vector:
$\hat O_c=\frac{1}{2}(I_1\otimes I_2+\gamma_1^0\otimes\gamma_2^0)$
which is perhaps the simplest choice from the more general ones [1-4]
and provides the existence of the stable discrete energy levels.
The one--gluon exchange part of the potential was neglected since we were
interested in the retardation corrections, coming from the confining part
of the potential. Further, for the simplicity, we have restricted ourselves
to the equal--mass case. We have used the Logunov--Tavkhelidze quasipotential
approach \cite{LT} in order to reduce the original, 4--dimensional Bethe--Salpeter
equation to the 3--dimensional one, which can be numerically solved with the use
of the conventional mathematical methods.

The quasipotential equation for the 3--dimensional equal--time wave function
$\tilde\varphi({\vec p})$ (in c.m.f.) is written in the following form \cite{KR,KRYAF}:
\eq
[M_B-h_1({\vec p})-h_2(-{\vec p})]\tilde\varphi({\vec p})=
-i\gamma_1^0\gamma_2^0~\frac{4}{3}\int\frac{d^3{\vec q}}{(2\pi)^3}
\tilde{\underline V}(M_B;{\vec p},{\vec q})\tilde\varphi({\vec q})
\en
\noindent
where $M_B$ is the mass of the bound \qq system and $h_i=\vec\alpha_i\vec p_i+m\gamma_i^0$,
$m$ being the mass of the constituent quark. Keeping in mind that the retardation
corrections for heavy quarkonia are expected to be small, further we restrict ourselves
to the first--order quasipotential formalism, where $\tilde{\underline V}$
is given by the following expression \cite{KR,KRYAF}:
\eq
\tilde{\underline V}^{(1)}(M_B;{\vec p},{\vec q})=
<{\vec p}|\tilde{\underline G_0}^{-1}\widetilde{G_0KG_0}\tilde{\underline G_0}^{-1}|{\vec q}>
\en

Here $G_0$ is the free two--fermion Green's function, K is the Bethe--Salpeter
equation kernel and the procedure $\tilde A$ for any operator $A$ is
defined as follows:
\eq
\tilde A(P;{\vec p},{\vec q})=\int\frac{dp_0}{2\pi}A(P;p,q)\frac{dq_0}{2\pi}
\en
\noindent
and
$$
\tilde G_0=\tilde{\underline G_0}\gamma_1^0\gamma_2^0\Pi ;~~
\Pi =(\Lambda_1^{(+)}\Lambda_2^{(+)}-\Lambda_1^{(-)}\Lambda_2^{(-)})\gamma_1^0\gamma_2^0;~~
$$
\eq
\Lambda_i^{(\pm)}=\frac{w_i\pm h_i}{2w_i};~~
w_i=(m^2+{\vec p}_i^2)^{\frac{1}{2}}
\en

The partial--wave decomposition of the eq. (9) is done with the use of the
standard technique \cite{ACKR,KR,KRYAF}. In the wave function $\tilde\varphi({\vec p})$
we retain only the contribution from the "double--positive" component
$\tilde\varphi^{(++)}({\vec p})=\Lambda_1^{(+)}\Lambda_2^{(+)}\tilde\varphi({\vec p})$,
as the role of the negative--energy component is shown to be small provided the
stable solutions exist \cite{ACKR}. Using the substitution
\eq
\tilde\varphi^{(++)}({\vec p})=
\left(\frac{w+m}{2w}\right)^{\frac{1}{2}}\pmatrix{1\cr \frac{ \vec\sigma_1\vec p}{w+m}}\otimes
\left(\frac{w+m}{2w}\right)^{\frac{1}{2}}\pmatrix{1\cr \frac{-\vec\sigma_2\vec p}{w+m}}
\tilde\chi^{(+)}({\vec p})
\en
\noindent
and decomposing the Pauli spinor $\tilde\chi^{(+)}({\vec p})$ into partial waves
\eq
\tilde\chi^{(+)}({\vec p})=\sum_{LSJM_J}<{\vec n}|LSJM_J>\tilde R^{(+)}_{LSJ}(p);~~
{\vec n}=\frac{{\vec p}}{p};~~S=0,1
\en
\noindent
from the eq. (9) we obtain
$$
[M_B-2w(p)]\tilde R^{(+)}_{LSJ}(p)=\frac{4}{3}\int_0^{\infty}q^2dq\left(
\tilde V^{(+)}_L(p,q)\frac{1}{2}\left(1+\frac{m^2}{w(p)w(q)}\right)+\right.
$$
\eq
\left. +(L-J)\frac{4J(J+1)}{(2J+1)^2}
(\tilde V^{(+)}_{J-1}(p,q)-\tilde V^{(+)}_{J+1}(p,q))
\frac{(w(p)-m)(w(q)-m)}{4w(p)w(q)}\right)\tilde R^{(+)}_{LSJ}(q)
\en
\noindent
where, for the case of local quasipotential
\eq
\tilde V^{(+)}_L(p,q)=\int r^2drj_L(pr)\tilde V_c(r;M_B)j_L(qr)
\en
\noindent
$j_L$ being the spherical Bessel function and we, as in ref. \cite{KR,KRYAF},
have neglected small mixing between $L=J-1$ and $L=J+1$ states.

After doing some algebra with the projection operators the positive--energy
component of the first--order quasipotential, corresponding to the interaction
kernel (4) with the use of eqs. (10), (11) can be written in the following
form:
$$
\tilde{\underline V}^{(+)}(M_B;{\vec p},{\vec q})=
\int d^3{\vec x}e^{-i({\vec p}-{\vec q}){\vec x}}
\frac{\Gamma\left(1+\frac{\alpha}{2}\right)}{\sqrt{\pi}\Gamma\left(\frac{1}{2}+\frac{\alpha}{2}\right)}r^{\alpha}\times
$$
\eq
\times\int_{-1}^{1}d\tau (1-\tau^2)^{(\alpha-1)/2}
(\theta(\tau)e^{i(M_B-w(p)-w(q))r\tau}+\theta(-\tau)e^{-i(M_B-w(p)-w(q))r\tau})
\en

Neglecting relativistic corrections in the exponentials
$M_B-w(p)-w(q)=M_B-2m+O(\frac{1}{m})=-\epsilon_B$, as well as the imaginary
part of this expression in analogy with the refs. \cite{KR,KRYAF,ARKH},
we arrive at the local first--order quasipotential:
$$
\tilde{\underline V}^{(+)}(r;\epsilon_B)=
\frac{\Gamma\left(1+\frac{\alpha}{2}\right)}{\sqrt{\pi}\Gamma\left(\frac{1}{2}+\frac{\alpha}{2}\right)}r^{\alpha}
\int_{-1}^{1}d\tau (1-\tau^2)^{(\alpha-1)/2}\cos\epsilon_Br\tau =
$$
\eq
=\left|\frac{2r}{\epsilon_B}\right|^{\alpha /2}\Gamma\left(1+\frac{\alpha}{2}\right)
J_{\alpha /2}(|\epsilon_Br|)
\en

Hence, the relativistic generalization of the static potential $V_c(r)=kr+c$
gives the first--order local quasipotential
\eq
\tilde V_c(r,\epsilon_B)=k\frac{\sin\epsilon_Br}{\epsilon_B}+cJ_0(\epsilon_Br)
\en
\noindent
which accounts for the retardation effect and reduces to $V_c(r)$ in the
limit $\epsilon_B\rightarrow 0$.

As it can be seen from eq. (19), the account for the dynamical retardation
effect in the case of pure linear potential effectively leads to the colour
screening at an intermediate distances. It should be pointed out that
such behaviour qualitatively agrees with the results of calculations
for the unquenched lattice fermions in QCD \cite{LAER}. At a larger distances
the deviation of the retarded potential from the static one becomes significant,
and one can no further rely on the first--order calculations. Note, however,
that for the case of heavy quarkonia the latter difficulty causes no trouble
since the wave function of the \qq bound system in this case rapidly vanishes
with the increase of $r$ and, therefore, does not "feel" the oscillating
"tail" of the potential at a large distances.

\section{Results and Discussion}

\par
It is not obvious from the beginning whether the potential (19) leads to
the discrete energy levels due to its oscillating behaviour at $r\rightarrow\infty$.
Let us, therefore consider the equation (15) with the potential (19) in detail.
Passing to the nonrelativistic limit and neglecting for a moment the "constant"
term in (19), proportional to $c$, in the configuration space we obtain the
following differential equation:

\eq
f''(z)+(a\cos 2z+b)f(z)=0
\en

\noindent
where $f(r)=rR(r)$, $R(r)\equiv R_0(r)$ being the radial wave function of the
bound state in the configuration space (for simplicity we assume the angular
momentum, $L=0$), $z=\frac{1}{2}((M_B-2m)r-\pi/2)$, $a=\frac{k(M_B-2m)}{3m^3}$,
$b=\frac{(M_B-2m)}{4m^3}$, and the boundary conditions imposed on $f(z)$
are $f\left( -\frac{\pi}{4}\right) =0$ and $f(+\infty )=0$.

The equation (20) has been extensively studied in the mathematical physics
(see, e.g. \cite{KAMKE}). We shall remind some results of this investigation.
Namely, if $f_1(z)$ is the particular solution of the eq. (20) with the following
initial conditions:
\eq
f_1(0)=1,~~~f'_1(0)=0
\en

\noindent
and
$$
\cosh 2\pi\mu=f_1(\pi )
$$

Then the general solution of the equation (20) has the form:

\begin{equation}
f(z)=\left\{
\begin{array}{l}
     C_1{\rm e}^{2\mu z}\varphi_1(z)+C_2{\rm e}^{-2\mu z}\varphi_2(z);~~\cosh 2\pi\mu >1 \\
(C_1\cos 2\nu z+C_2\sin 2\nu z)\varphi_1(z)+(C_2\cos 2\nu z-C_1\sin 2\nu z)\varphi_2(z);\\
~~~~~~~~~~~~~~~~~~~~~~~~~~~~~~~~~~~~~~~~~~~~~~~~~|\cosh 2\pi\mu |<1;~~\mu ={\rm i}\nu \\
C_1{\rm e}^{2\rho z}\varphi_1(z)+C_2{\rm e}^{-2\rho z}\varphi_2(z);~~\cosh 2\pi\mu <-1;~~\mu =\rho +\frac{{\rm i}}{2} \\
\end{array}
\right.
\end{equation}

\noindent

$\varphi_1(z)$ and $\varphi_2(z)$ being the periodic functions in $z$
with the period $\pi$.

Due to the fact that we consider the equation (20) on the semi--infinite
interval $-\frac{\pi}{4}\leq z<+\infty$, it is possible to find the normalizable
solutions decreasing exponentially at $z\rightarrow +\infty$ ($C_1=0$ and
$|\cosh 2\pi\mu |>1$, eq. (22)). The eigenvalue condition then reads as:

\eq
\varphi_2\left(-\frac{\pi}{4}\right) =0,~~~~|\cosh 2\pi\mu |>1
\en

Thus, the equation (20), despite the oscillating behaviour of the potential at the
spatial infinity, allows for the discrete spectrum provided $|\cosh 2\pi\mu |>1$,
corresponding to the condition $M_B-2m<0$ in the limit $|M_B-2m|<<2m$.
Adding the constant term, proportional to $c$, it is natural to suppose that,
for a small $|M_B-2m|$ the discrete energy levels exist for $M_B-2m-\frac{4}{3}c<0$.
Thus the potential (19) in the nonrelativistic limit acts like the potential
well. Note that the similar potential (the rising potential screened at large distances, $r>1~{\rm Fm}$
was successfully used for the description of meson spectrum in the framework
of coupled Dyson--Schwinger and Bethe--Salpeter equations e.g. in ref. \cite{MUNZ}.
Therefore we expect that the equation (15) with the potential (19) gives the reasonable
description of the low--lying meson states.

At the next step we have attempted to solve the eq. (15) numerically, expanding the unknown radial
wave function $\tilde R^{(+)}_{LSJ}(p)$ in the complete orthonormalized basis of the
nonrelativistic oscillator wave functions \cite{KOP,ACKR,KR,KRYAF}
\eq
\tilde R^{(+)}_{LSJ}(p)=p_0^{-3/2}\sum_{n=0}^{\infty}c^{(+)}_{nLSJ}R_{nL}(p/p_0)
\en
\noindent
where
\eq
R_{nL}(z)=
\left(\frac{2\Gamma\left( n+L+\frac{3}{2}\right)}{\Gamma (n+1)}
\right)^{\frac{1}{2}}
\frac{1}{\Gamma\left( L+\frac{3}{2}\right) }
z^L\exp\left( -\frac{1}{2}z^2\right) ~_1F_1\left( -n,L+\frac{3}{2},z^2\right)
\en
\noindent
and $p_0$ is an arbitrary scale parameter. Inserting (24) in the equation (15)
and truncating the sum at some fixed value $N_{max}$, we arrive at a system
of linear algebraic equations for the coefficients $c^{(+)}_{nLSJ}$.
If the procedure converges with the increase of $N_{max}$, the eigenvalues
$M_B$ are determined from this system of equations. The calculations show
that the final results do not depend on the scale parameter $p_0$, but the
appropriate choice of this parameter leads to the faster convergence of
the series (24). It should be stressed that if the solution of the equation
(15) does not exist, this reveals in the divergence of the procedure with the
increase of $N_{max}$ despite the fact that the potential matrix elements
are calculated in the exponentially damping wave function basis.

Since the potential (19) depends on the unknown binding energy, $\epsilon_B=2m-M_B$,
of the \qq system, the equation (15) is solved with the use of the iteration
method. Namely, we solve the equation with the static potential, $V_c(r)=kr+c$
and determine the eigenvalues $M_B^{(st)}$. At the next step these static values
are substituted into the potential (19) in order to determine the corrected
spectrum which, in its turn, is used, as an input, in the next iteration.
We have checked that, typically after 10--15 steps, the iteration procedure
converges for the most low--lying heavy quarkonia energy levels.

In the table 1 the results of calculations of the dynamical retardation
corrections to the heavy quarkonia mass spectrum are presented. In these
calculations the parameters $k$ and $c$ were taken to be $k=0.21~GeV^2$,
$c=-1.0~GeV$. The constituent quark masses were chosen to be $m_c=1.72~GeV$
and $m_b=5.10~GeV$ in order to fit $J/\psi$ and $\Upsilon$ masses.
As we see from table 1, this set of parameters gives the reasonable description of
heavy meson mass spectrum in the static approximation. As expected, the
dynamical retardation corrections turn out to be small (typically a few
percent) for all low--lying quarkonia states given in this table.

As we see, the present approach to the relativistic generalization of the
static confining potential does not suffer from some of difficulties
inherent of the conventional ones [5-10]. Namely, an additional mass scale
parameter does not appear in the "relativized" counterpart of the static
potential. Moreover, this approach possesses a smooth static limit
and yields small corrections due to the retardation in the heavy quarkonia
mass spectrum. This, unlike the results of refs. \cite{BRAM,KRYAF}, does not
manifestly contradict with our expectation at least in the heavy--quark sector.
An important remark, however, should be made, concerning the approach
suggested in the present paper. Namely, the kernels off the type considered here
(eqs. (4), (5)) can never be obtained in the Euclidean formulation of QCD
(see eq. (7)). Consequently, such fundamental properties, as microcausality
and unitarity, do not necessarily continue to hold in this approach. Of course,
one can appeal again to the 2--dimensional QCD, where, in spite of the
principal--value prescription used in the gluon propagator, the uncoloured current--current
correlation functions are known to have the correct analytical structure
\cite{BROW}. Note, however, that in 2 dimensions the gluon field is not
a dynamical one \cite{EIN}, unlike the case of 4 dimensions, where the different
components of the gluon field (dynamical and nondynamical) may be subject
to the different boundary conditions.
Here one finds an analogy with the result obtained in the ref. \cite{CHENGTSAI}
where, in particular, it was demonstrated that the ghost degrees of freedom
(in the covariant gauge) can be eliminated from the theory at the cost of
Feunman's boundary conditions in the propagator for timelike and longitudinal
gluons, being replaced by the principal--value prescription. An alternative
point of view was presented in ref. \cite{BLAHA}, where the quantization procedure
for the gauge field is carried out with the account of the fact that this field
is never observed in $in-$ or $out-$ states. As a result, the principal--value
prescription emerges in the propagator for all components of the gauge field instead
of the conventional Feunman one. Consequently, the anzats (4), corresponding
to the "minimal" relativistic generalization of the static interquark interactions,
can be regarded as an appropriate candidate for the truly confining interquark
interaction kernel in 4 dimensions.
\vspace{.5cm}

The authors are thankful to Dr. B.Magradze for many helpful discussions.
One of the authors (A.R.) acknowledges the financial support under grant
UR-94-6.7-2042.

\vspace{.7cm}

{\Large{\bf Appendix A}}
\vspace{.2cm}

Below we define the "average" of the static kernel (1) over all possible
directions of the unit vector $n$ in the 4--dimensional Minkowski space.
This approach can be applied to the case of any space--time dimensions and,
in particular, to the case of 2 dimensions, considered in the text.

We start from the Lorentz--invariant regularization of the "average" of
$\delta$-function, entering the expression (1):
\setcounter{equation}{0}
\eq
\renewcommand{\theequation}{A.\arabic{equation}}
<\delta(nx)>_n=\lim_{\epsilon\to 0}\int_{-\infty}^{\infty}\frac{d\lambda}{2\pi}
<e^{i\lambda(nx)-\epsilon\lambda^2}(\theta(x^2)+\theta(-x^2))>_n\equiv
\bar\delta_++\bar\delta_-
\en

Let us consider first the case $x^2>0$
\eq
\renewcommand{\theequation}{A.\arabic{equation}}
\bar\delta_+=\theta(x^2)(x^2)^{-1/2}\lim_{\epsilon\to 0}\int_{-\infty}^{\infty}
\frac{d\lambda}{2\pi}<e^{i\lambda(n\eta_+)-\epsilon\lambda^2}>_n;~~
\eta_+^{\mu}={x^{\mu}}/{\sqrt{x^2}};~~
\eta_+^2=1
\en

Expanding $e^{i\lambda(n\eta_+)}$ in powers of $\lambda$ and defining the
"average" in a familiar manner:
$$
<n_{\mu_1}\cdots n_{\mu_k}>_n=0~~~~~{\rm for}~k~{\rm odd}
$$
\eq
\renewcommand{\theequation}{A.\arabic{equation}}
<1>_n=1;~~~<n_{\mu_1}n_{\mu_2}>_n=\frac{1}{4}g_{\mu_1\mu_2};~~~etc
\en
\noindent
we obtain
\eq
\renewcommand{\theequation}{A.\arabic{equation}}
<e^{i\lambda(n\eta_+)}>_n=1+\frac{(i\lambda)^2}{2!}\frac{1}{4}+\cdots =F_+(\lambda)
\en

In order to calculate $F_+(\lambda)$ we consider the 4--dimensional Euclidean space
and define the average:
\eq
\renewcommand{\theequation}{A.\arabic{equation}}
<e^{i\lambda(n^E\eta_+^E)}>_{n^E}=\frac{1}{\pi^2}\int d^4n^E\delta({n^E}^2-1)
e^{i\lambda(n^E\eta_+^E)};~~~{n^E}^2={\eta_+^E}^2=1
\en

Using the relations
$$
<n^E_{\mu_1}\cdots n^E_{\mu_k}>_{n^E}=0~~~~~{\rm for}~k~{\rm odd}
$$
\eq
\renewcommand{\theequation}{A.\arabic{equation}}
<1>_{n^E}=1;~~~<n^E_{\mu_1}n^E_{\mu_2}>_{n^E}=\frac{1}{4}\delta_{\mu_1\mu_2};~~~etc
\en
\noindent
it is easy to ensure that (A.5) coincides with $F_+(\lambda)$. Passing in the integral
in (A.5) to the angular variables and substituting the result into (A.2),
we finally obtain
\eq
\renewcommand{\theequation}{A.\arabic{equation}}
\bar\delta_+=\theta(x^2)(x^2)^{-1/2}\lim_{\epsilon\to 0}\frac{1}{4\pi^2}
\int_{-\infty}^{\infty}d\lambda\int_{-1}^{1}dy\sqrt{1-y^2}e^{i\lambda y-\epsilon\lambda^2}=
\theta(x^2)(x^2)^{-1/2}\frac{1}{2\pi}
\en

The case $x^2<0$ can be considered in a similar way. However, due to the
change of sign $\eta_-^2=-1$, where $\eta_-^{\mu}={x^{\mu}}/{\sqrt{-x^2}}$,
in the integrand $e^{i\lambda y}$ is replaced by $e^{\lambda y}$
\eq
\renewcommand{\theequation}{A.\arabic{equation}}
\bar\delta_-=\theta(-x^2)(-x^2)^{-1/2}\lim_{\epsilon\to 0}\frac{1}{4\pi^2}
\int_{-\infty}^{\infty}d\lambda\int_{-1}^{1}dy\sqrt{1-y^2}e^{\lambda y-\epsilon\lambda^2}
\en
\noindent
and, as a result, $\bar\delta_-$ diverges in the limit $\epsilon\rightarrow 0$.
Including this divergent (Lorentz--invariant) factor in the overall normalization
constant ${\cal N}$, we see, that only the term
$\bar\delta_-$ survives in the limit $\epsilon\rightarrow 0$.
Finally, substituting (A.8) into the eq. (2) and determining the normalization
constant in the static limit, we again arrive at the expression (4)
for the relativistic Bethe--Salpeter kernel.

\newpage
\bc
\begin{tabular}{|l|c|c|c|c|c|c|}
\hline
Mesons                   & $J^{PC}$ & $N^{2S+1}L_J$ & 1      & 2      & 3     & 4                        \\
\hline
$c\bar c$                &          &               &        &        &       &                          \\
$\eta_c(2.980)$          & $0^{-+}$ & $1^1S_0$      & 3.095  & 3.204  & 0.109 & 3.5$\cdot {\rm 10}^{-2}$ \\
$\eta'_c(3.590)$         & $0^{-+}$ & $2^1S_0$      & 3.690  & 3.739  & 0.049 & 1.3$\cdot {\rm 10}^{-2}$ \\
$J/\psi (3.097)$         & $1^{--}$ & $1^3S_1$      & 3.096  & 3.204  & 0.108 & 3.5$\cdot {\rm 10}^{-2}$ \\
$\psi'(3.686)$           & $1^{--}$ & $2^3S_1$      & 3.691  & 3.742  & 0.051 & 1.3$\cdot {\rm 10}^{-2}$ \\
$h_{c1}(3.526)$          & $1^{+-}$ & $1^1P_1$      & 3.445  & 3.460  & 0.014 & 4.2$\cdot {\rm 10}^{-3}$ \\
$\chi_{c0}(3.414)$       & $0^{++}$ & $1^3P_0$      & 3.445  & 3.460  & 0.014 & 4.2$\cdot {\rm 10}^{-3}$ \\
$\chi_{c1}(3.511)$       & $1^{++}$ & $1^3P_1$      & 3.445  & 3.460  & 0.014 & 4.2$\cdot {\rm 10}^{-3}$ \\
$\chi_{c2}(3.556)$       & $2^{++}$ & $1^3P_2$      & 3.445  & 3.461  & 0.014 & 4.2$\cdot {\rm 10}^{-3}$ \\
$b\bar b$                &          &               &        &        &       &                          \\
$\eta_b$                 & $0^{-+}$ & $1^1S_0$      & 9.463  & 9.619  & 0.156 & 1.7$\cdot {\rm 10}^{-2}$ \\
$\eta'_b$                & $0^{-+}$ & $2^1S_0$      & 9.899  & 9.966  & 0.067 & 6.8$\cdot {\rm 10}^{-3}$ \\
$\Upsilon(9.460)$        & $1^{--}$ & $1^3S_1$      & 9.463  & 9.619  & 0.156 & 1.7$\cdot {\rm 10}^{-2}$ \\
$\Upsilon'(10.023)$      & $1^{--}$ & $2^3S_1$      & 9.899  & 9.966  & 0.067 & 6.8$\cdot {\rm 10}^{-3}$ \\
$\Upsilon''$             & $1^{--}$ & $1^3D_1$      & 9.938  & 9.993  & 0.055 & 5.5$\cdot {\rm 10}^{-3}$ \\
$\Upsilon'''(10.355)$    & $1^{--}$ & $3^3S_1$      & 10.250 & 10.255 & 0.005 & 5.1$\cdot {\rm 10}^{-4}$ \\
$\Upsilon^{IV}$          & $1^{--}$ & $2^3D_1$      & 10.277 & 10.291 & 0.014 & 1.4$\cdot {\rm 10}^{-3}$ \\
$h_{b1}$                 & $1^{+-}$ & $1^1P_1$      & 9.720  & 9.835  & 0.115 & 1.2$\cdot {\rm 10}^{-2}$ \\
$\chi_{b0}(9.860)$       & $0^{++}$ & $1^3P_0$      & 9.720  & 9.835  & 0.115 & 1.2$\cdot {\rm 10}^{-2}$ \\
$\chi_{b1}(9.892)$       & $1^{++}$ & $1^3P_1$      & 9.720  & 9.835  & 0.115 & 1.2$\cdot {\rm 10}^{-2}$ \\
$\chi_{b2}(9.913)$       & $2^{++}$ & $1^3P_2$      & 9.720  & 9.835  & 0.115 & 1.2$\cdot {\rm 10}^{-2}$ \\
$h'_{b1}$                & $1^{+-}$ & $2^1P_1$      & 10.097 & 10.109 & 0.013 & 1.3$\cdot {\rm 10}^{-3}$ \\
$\chi'_{b0}(10.232)$     & $0^{++}$ & $2^3P_0$      & 10.097 & 10.109 & 0.013 & 1.3$\cdot {\rm 10}^{-3}$ \\
$\chi'_{b1}(10.255)$     & $1^{++}$ & $2^3P_1$      & 10.097 & 10.109 & 0.013 & 1.3$\cdot {\rm 10}^{-3}$ \\
$\chi'_{b2}(10.268)$     & $2^{++}$ & $2^3P_2$      & 10.097 & 10.109 & 0.013 & 1.3$\cdot {\rm 10}^{-3}$ \\
\hline
\end{tabular}
\ec

\bc
Table 1. The dynamical retardation corrections to the heavy quarkonia mass spectrum.
1) The meson mass in the static approximation, $M_B^{(st)}$ (GeV),
2) The meson mass with an account of the retardation effect, $M_B^{(ret)}$ (Gev),
3) The size of the retardation correction, $M_B^{(ret)}-M_B^{(st)}$ (GeV),
4) $|M_B^{(ret)}-M_B^{(st)}|/M_B^{(st)}$.
\ec

\end{document}